\documentclass[aip,reprint,superscriptaddress,jap]{revtex4-1}

\usepackage{booktabs}
\usepackage{multirow}
\usepackage{amsmath}
\usepackage{graphicx} 

\newcommand{\bournonite}{CuPbSbS$_3$}
\newcommand{\enargite}{Cu$_3$AsS$_4$}

\draft 

\begin{document}


\title{Finding a junction partner for candidate solar cell absorbers enargite and bournonite from electronic band and lattice matching}



\author{Suzanne K. Wallace}
\affiliation{Department of Chemistry, Centre for Sustainable Chemical Technologies, University of Bath, Claverton Down, Bath, BA2 7AY, UK}
\affiliation{Department of Materials, Imperial College London, Exhibition Road, London SW7 2AZ, UK}

\author{Keith T. Butler}
\affiliation{ISIS Neutron and Muon Source, Rutherford Appleton Laboratories, Didcot, Oxfordshire, OX11 0QX, UK}

\author{Yoyo Hinuma}
\affiliation{Center for Frontier Science, Chiba University, Chiba 263-8522, Japan}
\affiliation{Center for Materials Research by Information Integration, Research and Services Division of Materials Data and Integrated System, National Institute for Materials Science, Tsukuba 305-0047, Japan}

\author{Aron Walsh}
\affiliation{Department of Materials, Imperial College London, Exhibition Road, London SW7 2AZ, UK}
\affiliation{Department of Materials Science and Engineering, Yonsei University, Seoul 03722, Korea}
\email[]{a.walsh@imperial.ac.uk}

\date{\today}

\begin{abstract}
An essential step in the development of a new photovoltaic (PV) technology is choosing appropriate electron and hole extraction layers to make an efficient device. We recently proposed the minerals enargite (\enargite) and bournonite (\bournonite) as materials that are chemically stable with desirable optoelectronic properties for use as the absorber layer in a thin-film PV device. For these compounds, spontaneous lattice polarization with internal electric fields --- and potential ferroelectricity --- may allow for enhanced carrier separation and novel photophysical effects. In this work, we calculate the ionization potentials for non-polar surface terminations and propose suitable partners for forming solar cell heterojunctions by matching the electronic band edges to a set of candidate electrical contact materials. We then further screen these candidates by matching the lattice constants and identify those that are likely to minimise strain and achieve epitaxy. This two-step screening procedure identified a range of unconventional candidate contact materials including SnS$_2$, ZnTe, WO$_3$, and Bi$_2$O$_3$. 
\end{abstract}


\maketitle 

\section{Introduction}
Solar power is an attractive source of sustainable electricity. 
Technological breakthroughs to enable high-efficiency photovoltaic (PV) devices without the need to use scarce material components and with low manufacturing costs would secure solar power as a future power source. 
Exploiting non-centrosymmetry and lattice polarization in `photoferroic' materials could provide new pathways to high-efficiency PV devices. Phenomena referred to as `anomalous' and `bulk' PV effects in polar materials have demonstrated photovoltages orders of magnitude greater than the optical band gap and photocurrents in bulk, single-crystal absorbers in the absence of a typical p-n junction for carrier separation \cite{Anomalous_PV, Bulk_PV, keith_FE-PV} . 
On-going research efforts are exploring the theory behind these observed phenomena \cite{Rappe, BPE_transport, Rappe2, Kirchartz_review}. 

We recently identified three naturally-ocurring minerals as candidate photoferroic materials based on their optical band gaps and polar crystal structures \cite{sulfosalts_paper1}, including enargite (\enargite) and bournonite (\bournonite) . To our knowledge, to date only one study has made solar cells out of any of these materials. 
In Ref. \citenum{enargite_SC}, solar cells were made from solution processed enargite using a device architecture developed for Cu(In,Ga)Se$_2$ (CIGS) solar cell technology. In this study the authors list non-optimal band alignment of the absorber layer with the device architecture as a likely limitation of the current solar cell performance. 
This is also the case for CuSbS$_2$ solar cells using device architecture optimised for CIGS absorber layers  \cite{CAS_alignment}. 
Mature technologies, such as CIGS-based devices, are still being optimised through improved band alignment with the n-type buffer layer  \cite{CIGS_new_buffer}
and non-optimal band alignment is also being considered as a limiting factor on the performance of Cu$_2$ZnSn(S,Se)$_4$-based solar cells \cite{Crovetto_review}. 
The optimization of device architecture for a new solar cell technology is a challenging and time-consuming process.

\begin{figure*}[]
\centering
  \includegraphics[width=16cm]{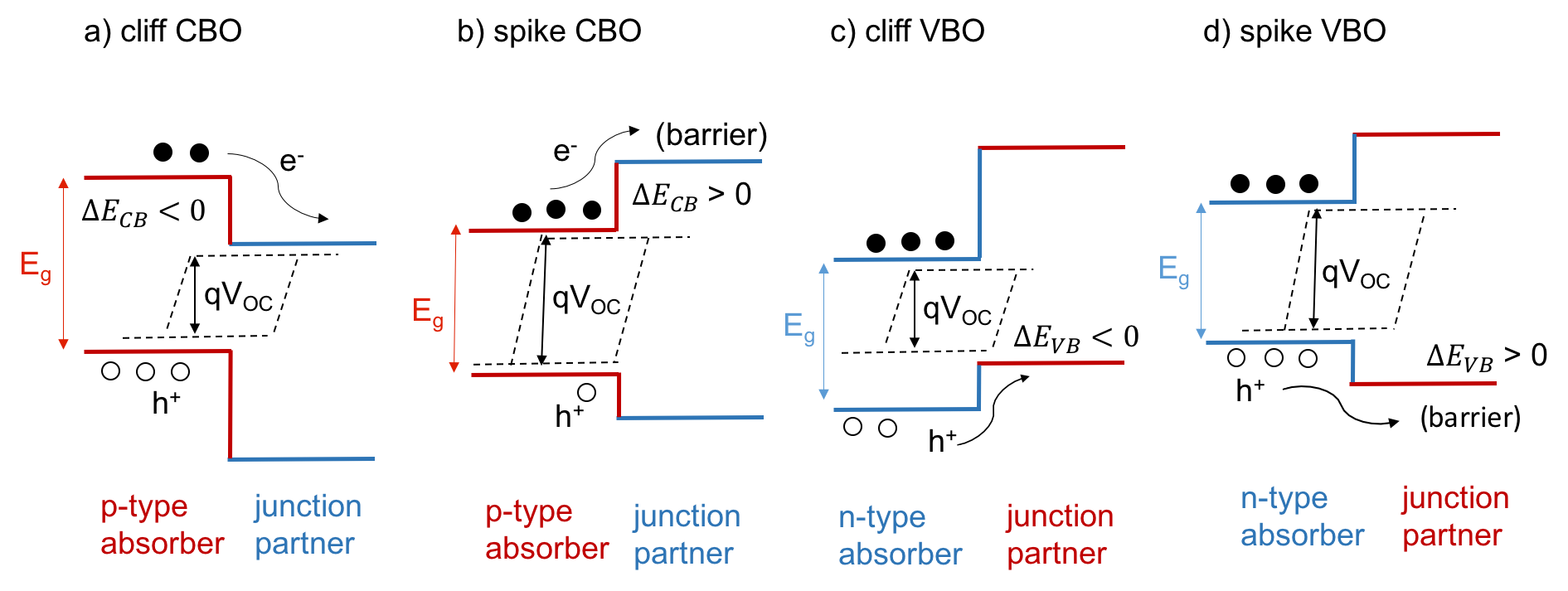}
  \caption{Illustration of four possible junctions present in a solar cell. 
  a)  A type II heterojunction between a p-type absorber and an n-type contact material where photoelectrons flow across the interface with no barrier at the negative `cliff' conduction band offset (CBO). The open-circuit voltage is reduced relative to the absorber band gap if the cliff is large.
  b) A type I heterojunction between a p-type absorber layer and an n-type contact material with a small positive (`spike') CBO. If the spike is suifficiently small, electrons can to tunnel through the barrier and across the interface. Holes are repelled from the interface region.
  c)  A type II heterojunction between an n-type absorber and a p-type contact material where photoelectrons flow across the interface with no barrier at the negative `cliff' valence band offset (VBO).  The open-circuit voltage is reduced relative to the absorber band gap if the cliff is large.
  d) A type I heterojunction between an n-type absorber layer and a p-type contact material with a small positive (`spike') VBO. If the spike is too large this barrier will impede transport across the interface.
  The schematics follow those presented in Ref. \citenum{CdTe_spike} and \citenum{Elisabetta}.}
  \label{junctions}
\end{figure*}

For many of the materials being studied for use as absorber layers in thin-film solar cells, such as chalcogenide semiconductors, it is not possible, or is very difficult, to achieve ambipolar doping. Therefore to achieve a p-n junction for many thin-film PV devices it is necessary to form an interface between materials with different optical band gaps, lattice constants and even crystal structures \cite{band_alignment_review}. 
In the extreme case when the two materials are poorly matched, differences in lattice constant and crystal structure at a heterojunction interface can introduce a large strain, resulting in poor epitaxy \cite{Keith_ELS}. Even for less extreme differences, small lattice mismatch at an interface generally introduces intra band gap defect states, which enhances Shockley-Read-Hall recombination, increasing dark currents, and reducing the open-circuit voltage of the device \cite{Nelson_CH8}.

In this work, we aim to accelerate the optimisation of solar cell device architectures for enargite (\enargite) and bournonite (\bournonite) by screening for candidate junction partners that could have optimal electronic band offsets and crystal lattices well-matched to minimise strain at the interface. The principles behind our screening criteria for optimal band offsets are outlined in the next section. Where there is no literature consensus as to whether the material is likely to be more easily doped p-type or n-type we screen for candidate contacts based on the relevant band offset for forming a solar cell junction for both cases.

\section{Band offsets for solar cell heterojunctions}

The band alignment at a solar cell junction is crucial to facilitate the separation of photo-excited electrons and holes to allow for extraction of the charge carriers before recombination can occur \cite{Heeger2013}. 
Although internal electric fields in the materials in this study may allow for a bulk photovoltaic effect for thin films (where the electrical asymmetry at a junction is not required for a photocurrent to be generated), a heterojunction would provide a global driving force for carrier collection at electrodes, while internal electric fields from the polar crystal structure could enhance carrier separation locally. 

Semiconductor junctions are classified as type I, II or III based on the band alignment; however only type I and type II are of interest for PV applications. 
A type II `staggered' junction can also be referred to as a `cliff-like' offset and a type I `straddling' junction can also be referred to as a `spike-like' offset, as illustrated schematically in Fig \ref{junctions}. 
For a p-type absorber layer, the minority carriers are electrons promoted into the conduction band (CB) of the absorber.
Therefore, the transport of electrons from the CB of the p-type absorber to the junction partner is important for determining device performance. The parameter of interest here is the conduction band offset (CBO) between the two materials.
However, for n-type absorbers it is photoexcited holes that are the minority carriers and so it is the magnitude of the valance band offset (VBO) between the n-type absorber and the junction partner that is important for charge extraction.

\begin{figure*}[]
\centering
  \includegraphics[width=16cm]{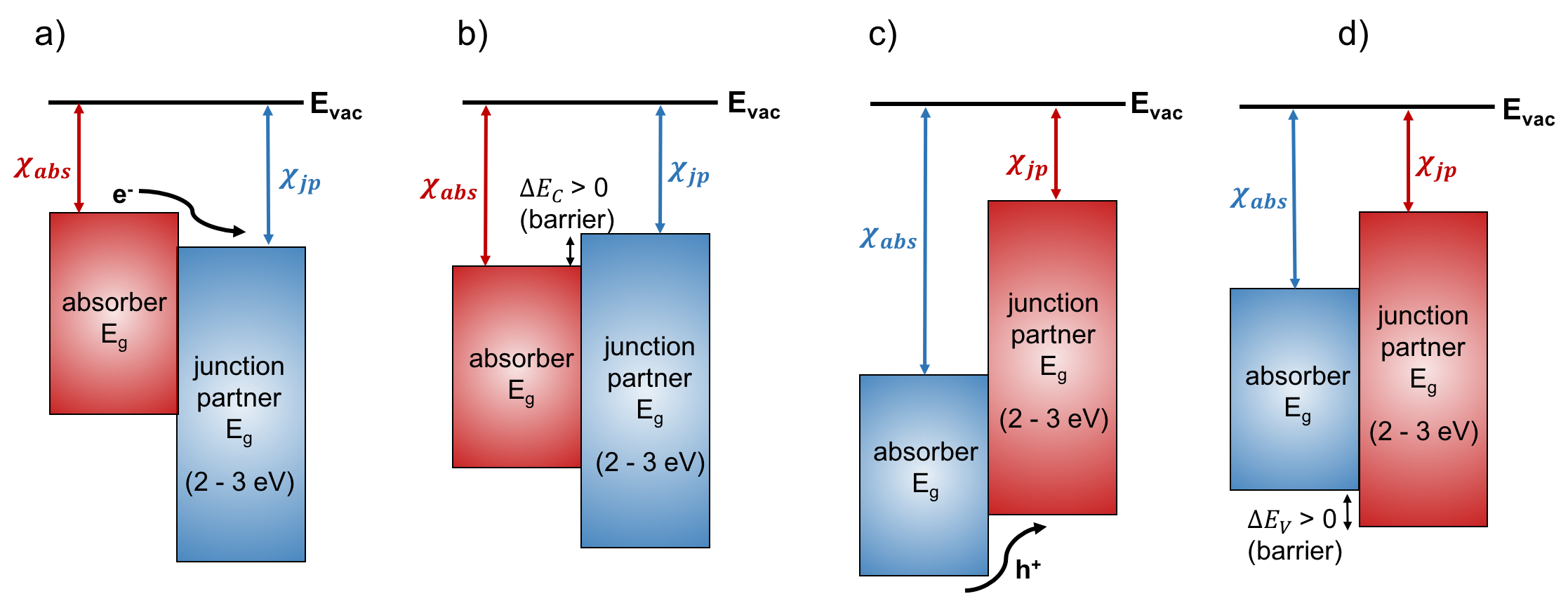}
  \caption{Heterojunction alignments for solar cells with respect to the external vacuum level, $E_{vac}$, based on the electron affinity (conduction band energy), $\chi_{abs}$. 
  a) Type II `staggered' offset with a p-type absorber and no barrier to minority photocarrier transport across the junction.
  b)  Type I `straddling' offset with a p-type absorber and a small barrier (postive $\Delta E_c$) to minority carrier transport. 
  c) Type II `staggered' offset with an n-type absorber and no barrier to minority photocarrier transport. 
  d) Type I `straddling' offset with an n-type absorber and a small barrier (postive $\Delta E_v$) to minority carrier transport. 
  The schematics follow those presented in Refs. \citenum{Hoffman-Wager} and \citenum{heterojunction_slides}.}
  \label{junctions_EA_rule}
\end{figure*}

It has been observed that for the p-type absorbers CIGS and CdTe, a spike offset (CBO within the range 0.1--0.3 eV), as shown in Fig. \ref{junctions}a, gives better device performance \cite{CdTe_spike, p-type_spike}.
Here, a small positive spike CBO creates an absorber inversion layer, resulting in a large hole barrier at the interface\cite{CdTe_spike}. 
The modest barrier to electron transfer means that electrons may still tunnel across the interface and be collected\cite{Nelson_CH8}. 
Electron-hole recombination at an interface with high defect densities is then thought to be suppressed due to an insufficient hole supply. 
In contrast, for a cliff offset where the CBO is negative, as shown in Fig \ref{junctions}b, there may be high concentrations of holes in the vicinity of the interface to assist interface recombination, thereby reducing the open-circuit voltage \cite{CdTe_spike}.

For the n-type absorber ZnSnN$_2$ the opposite trend has been observed, where a spike VBO is expected to give a poorer performance \cite{Elisabetta}. 
Here, a spike offset limits transport across the interface due to the larger effective mass of minority-carrier holes (compared to CIGS and CdTe \cite{CdTe_m*}) and associated lower hole mobility \cite{Elisabetta}. 
For ZnSnN$_2$, a small cliff is thought to be optimal.
 In this study, we use calculated effective masses to inform our choice of optimal band offsets for forming solar cell heterojunctions.

\section{Methodology}

\subsection{Electronic band and lattice matching}\label{matching_methods}
Screening for candidate heterojunction partners based on electronic band offsets and minimum lattice strain is conducted using the methodology and dataset of tabulated ionization potentials (IPs) and electron affinities (EAs) for candidate junction partners in Ref. \citenum{Keith_ELS} and ElectronLatticeMatch libraries \cite{ELS}. This dataset currently contains the electronic band gaps, IPs and EAs of 173 candidate heterojunction partners obtained either from experimental measurements or electronic structure calculations.

For p-type (n-type) absorbers we screen for candidate junction partners based on the CBO (VBO).
We look for a small cliff offset by selecting a band offset in the range 0 to -0.3 eV. 
However for cases where the minority carrier effective mass calculated in Ref. \citenum{sulfosalts_paper1} is less than 0.5$m_e$ we also look for a small spike offset in the range +0.1 to +0.3 eV. 
For CIGS +0.2 eV has been reported to be optimal \cite{p-type_spike} and +0.3 eV for CdTe \cite{CdTe_spike}.

We further limit our search to candidate junction partners where an interface with lattice strain less than 4\% is obtained.
We consider no defect states at the interface and allow no chemical intermixing, which is known to be present to a large extent at the CdTe:CdS contact \cite{CdTe_mixing,Ji-Sang_CdTe_intermixing}. 
For final candidate contacts we estimate the likely extent of interface intermixing based on the chemical similarities of the components of the two materials forming the heterojunction.



\subsection{Band alignment}

The alignment of the valence band energy to a common vacuum level, i.e. the ionization potential (IP),  can be peformed using techniques such as photoelectron spectroscopy or Kelvin probe microscopy and can be computed using first-principles calculations of surface slab models \cite{macro_pot_IP}.
The electron affinity ($\chi$) is the conduction band energy with respect to the vacuum level, which can be obtained by adding the value of the electronic band gap onto the IP of a material. 
Results from two decades of photoelectron spectroscopy experiments on CdTe and CIGS thin-film solar cells have been compared to density functional theory (DFT) calculations \cite{band_alignment_review}, where it was found that the energy band alignments for many interfaces were in good agreement. 
Theoretically predicted band alignments are usually the `intrinsic' or `natural' alignment for a particular combination of materials forming an interface \cite{band_alignment_review}, i.e. in the absence of defects,  interfacial reconstructions, or thermal effects.\cite{Bartomeu_T_effects}
This ideal band alignment therefore acts as a starting point to limit the search space for suitable junction partners.

The model used to predict the energy band alignment at solar cell heterojunctions in this study is the electron affinity rule (also known as Anderson's rule) where energies are aligned through the vacuum level \cite{Anderson_rule, keith_contacts}. 
The vacuum level of the two heterojunctions either side of the heterojunction are aligned to the same energy, the difference between the distance between the CBM and the vacuum ($\chi$) of each material is used to predict the CBO, as shown in Eq. \ref{CBO_calc}. 

We take semiconductor 1 to be the absorber with a band gap ($E_{g,abs}$) within the visible range (approximately 1.1--1.7 eV) and semiconductor 2 to be the transparent junction partner with a wider $E_{g, jp}$ in the range of 2--3 eV. The conduction band offset is defined as:
\begin{equation}\label{CBO_calc}
\Delta E_c = \chi_{jp} - \chi_{abs}
\end{equation}
Similarly, the valence band offset is determined through the difference in the ionization potentials ($IP = \chi + E_g$):
\begin{equation}\label{VBO_calc}
\begin{aligned}
\Delta E_v &  = IP_{abs} - IP_{jp}
\end{aligned}
\end{equation}
A negative $\Delta E_c$ or $\Delta E_v$ corresponds to a cliff CBO or cliff VBO respectively, this is then a `staggered gap'. These different cases are illustrated in Fig. \ref{junctions_EA_rule}.

\subsection{Computational details }
\subsubsection{Surface slab models }
Band energies are dependent upon the surface terminations of a crystal.
We therefore construct slab models for all possible non-polar surface terminations of the materials using the algorithm described in Ref. \citenum{Yoyo}. 
Symmetric slab models are then cut from relaxed unit cells; visualisations of the slab structures are given in the SI.

For the relaxation of the ion positions and volume of the unit cells, calculations are performed in VASP \cite{VASP1, VASP2} using the PBEsol functional \cite{PBEsol}, projector augmented wave core potentials\cite{VASP_PAW}, without including spin-orbit coupling (SOC) and with symmetry fixed until forces on the atoms are converged to within 0.005 eV per \AA. 
A plane wave cutoff energy of 350 eV is used and \textit{k}-grid densities of $6\times6\times6$ and $4\times4\times4$ were used to sample the electronic Brillouin zone for a 16 atom unit cell of enargite ({\enargite}) and a 24 atom unit cell of bournonite ({\bournonite}), repectively. 

To inform our later discussion for which candidate junction partners are likely to be the most important for devices, we calculate the surface energies for the slab models using Eq. \ref{surface_E} where $E_{bulk}$ is the total energy of the bulk crystal per formula unit, $n$ is the number of formula units in the surface slab and $A$ is the area of surface, of which there are two per slab model.
\begin{equation}\label{surface_E}
\gamma = \frac{E_{surf} - n E_{bulk}}{2A}
\end{equation}

\subsubsection{Ionization potential and electron affinity}
Calculations for planar averaged electrostatic potential of the slab models are also performed in VASP but using the HSE06 functional \cite{hse} with SOC and a single \textit{k}-point is used to sample the slab along the finite dimension. To calculate the ionization potentials of the surface slab models a macroscopic average technique is used. \cite{macro_pot_IP}
The difference between the macroscopic average of the vacuum potential and the bulk-like region of the surface slab is used to obtain the surface dipole shift, $D_s$. 
The ionization potential is then calculated using the eigenvalue of the valence band maximum ($\epsilon_{VBM}$) for the bulk crystal:
\begin{equation}\label{IP}
IP = D_s - \epsilon_{VBM}
\end{equation}
Electron affinities, $\chi$, are then calculated from the IP using the electronic band gap calculated with the HSE06 functional.

\section{Results and discussion}

\subsection{Electronic matching of junction partners}

The band energies calculated for each material are summarised in Table \ref{elec_offsets}.
We find a notable variation in the IP for different surface terminations of the same material, which may have implications for the simplicity of the junction fabrication, especially as the low energy surface terminations have significantly different IPs. 
From inspection of the slab geometries (shown in the SI),  slab models with larger IPs have denser surface structures with anion-rich terminations that could be associated with a  larger double layer.\cite{Bardeen_double_layer}

\begin{table}[]
\begin{ruledtabular}
\centering
\caption{Calculated ionization potentials (IP), electron affinities ($\chi$), DFT/HSE06 band gaps ($E_g$) all in units of eV and unrelaxed surface energies ($\gamma$) in units of eV per \AA$^2$  for symmetric and non-polar slab models of enargite ({\enargite}) and bournonite ({\bournonite}). The labels a and b refer to different surface cuts along the same [\textit{hkl}].}
\label{elec_offsets}
\begin{tabular}{@{}llllll@{}}
\toprule
Absorber                                                                      & Termination & E$_g$ & IP & $\chi$ & $\gamma$\\ \hline
\multirow{6}{*}     & (100)       & 1.32       & 4.97    & 3.64  & 0.050  \\
                                                                    Enargite                & (010)a     & 1.32       & 5.21    & 3.89  & 0.025  \\
                                                                       ({\enargite})              & (010)b     & 1.32       &   6.23      &   4.91   & 0.120   \\
                                                                                    & (110)     & 1.32       & 4.95    & 3.63  & 0.070  \\ \hline
\multirow{6}{*}{\begin{tabular}[c]{@{}l@{}}Bournonite \\ ({\bournonite})\end{tabular}} & (100)       & 1.68       & 5.61    & 3.93  & 0.042  \\
                                                                                    & (010)a     & 1.68       &    5.20     &     3.52   & 0.150 \\
                                                                                    & (010)b     & 1.68       &    5.21     &     3.53   & 0.106 \\
                                                                                    & (110)a     & 1.68       &   6.50      &    4.82    & 0.090 \\
                                                                                    & (110)b     & 1.68       &    6.04     &     4.36   & 0.037 \\ \cmidrule(l){1-6} 
\end{tabular}
\end{ruledtabular}
\end{table}

Experimental data for enargite suggests that the material exhibits native p-type conductivity \cite{enargite1995, sulfide_minerals_new}. 
For bournonite, there is variation in the literature between experimental measurements on natural samples measuring n-type conductivity \cite{sulfide_minerals_new} and theoretical prediction of the defect physics suggesting the material will be intrinsically p-type and difficult to dope n-type \cite{bournonite_defects}. 
Junction partners for enargite are screened based on the CBO only, while for bournonite the CBO and VBO are considered in turn to provide options for an p-type or n-type absorber layer.
We screen for both cliff and spike CBO for p-type bournonite, but only for a cliff VBO for n-type bournonite due to calculated heavy hole mass ($m_h >0.5 m_e$)\cite{sulfosalts_paper1} in line with our earlier discussion. 

\subsection{Low strain junctions}

\begin{figure}[h!]
\centering
  \includegraphics[width=0.5\textwidth]{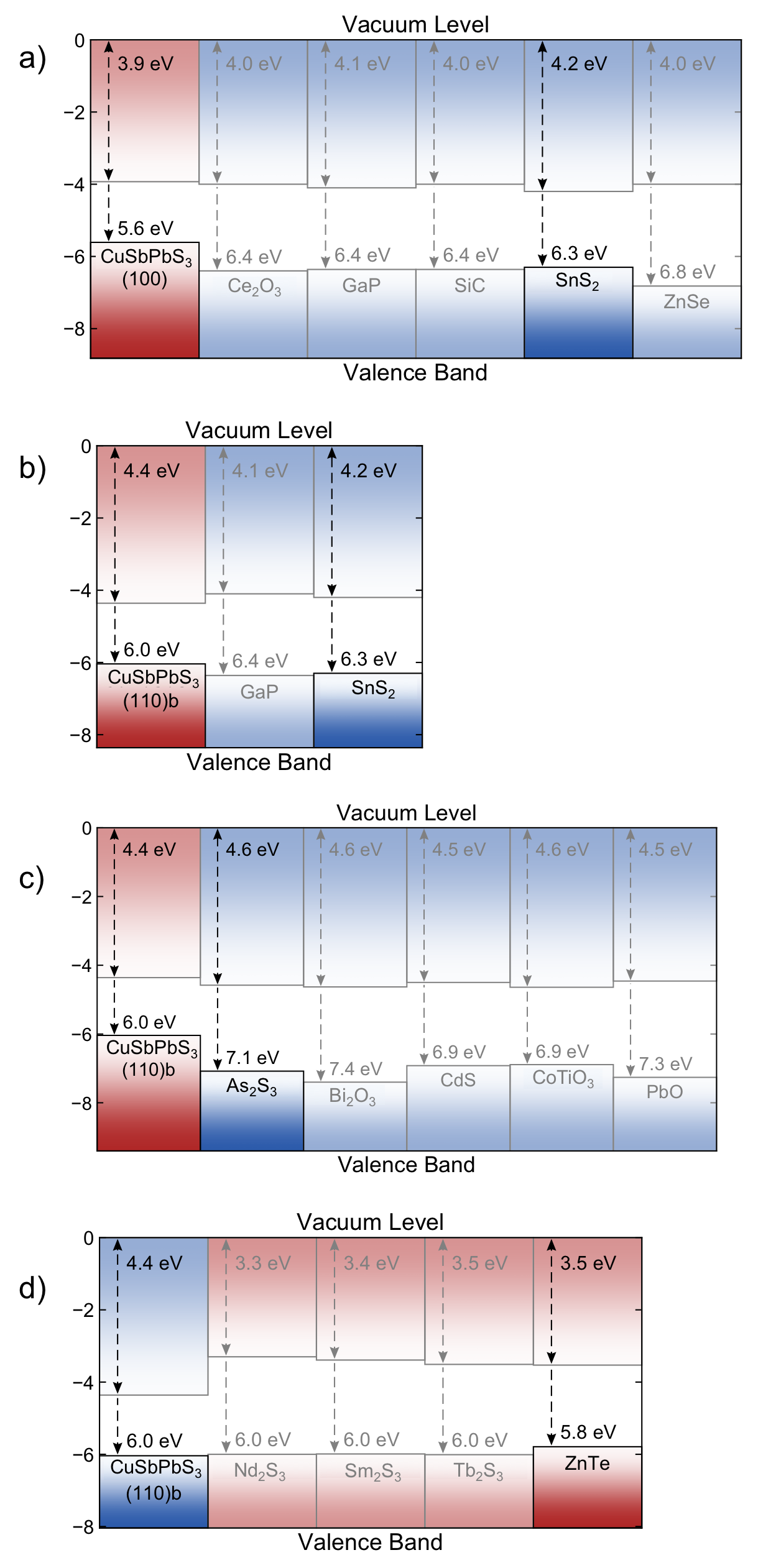}
  \caption{Candidate junction partners for p-type bournonite ({\bournonite}) absorber layer termination (100) for a) a cliff conduction band offset (CBO), b) (110)b termination for a spike CBO, c) cliff CBO and d) cliff valence band offset for n-type bournonite absorber. Low strain strain junction partner is shown in bold. Band alignment plots produced using the bapt package \cite{bapt}.}
  \label{b_110b_CBOspike+CBOcliff+VBOcliff}
\end{figure}

Candidates passing the electronic matching stage were screened for those with less than 4\%  lattice strain.
Candidates containing Fe were not considered further as Fe is often associated with fast non-radiative recombination
in solar cells due to possible excitations and recombination channels from a half occupied \textit{d}-shell (d$^5$ for Fe$^{3+}$ and d$^6$ for Fe$^{2+}$).
Where multiple structure files for the candidates were available on the Materials Project database \cite{materials_project}, unit cells for the most stable structures were preferentially selected for this lattice matching step, based on energy above the thermodynamic convex hull. 
All candidates considered in this second screening stage, including the materials project ID for the corresponding structure file, can be found in the SI.

\begin{figure}[h!]
\centering
  \includegraphics[width=0.5\textwidth]{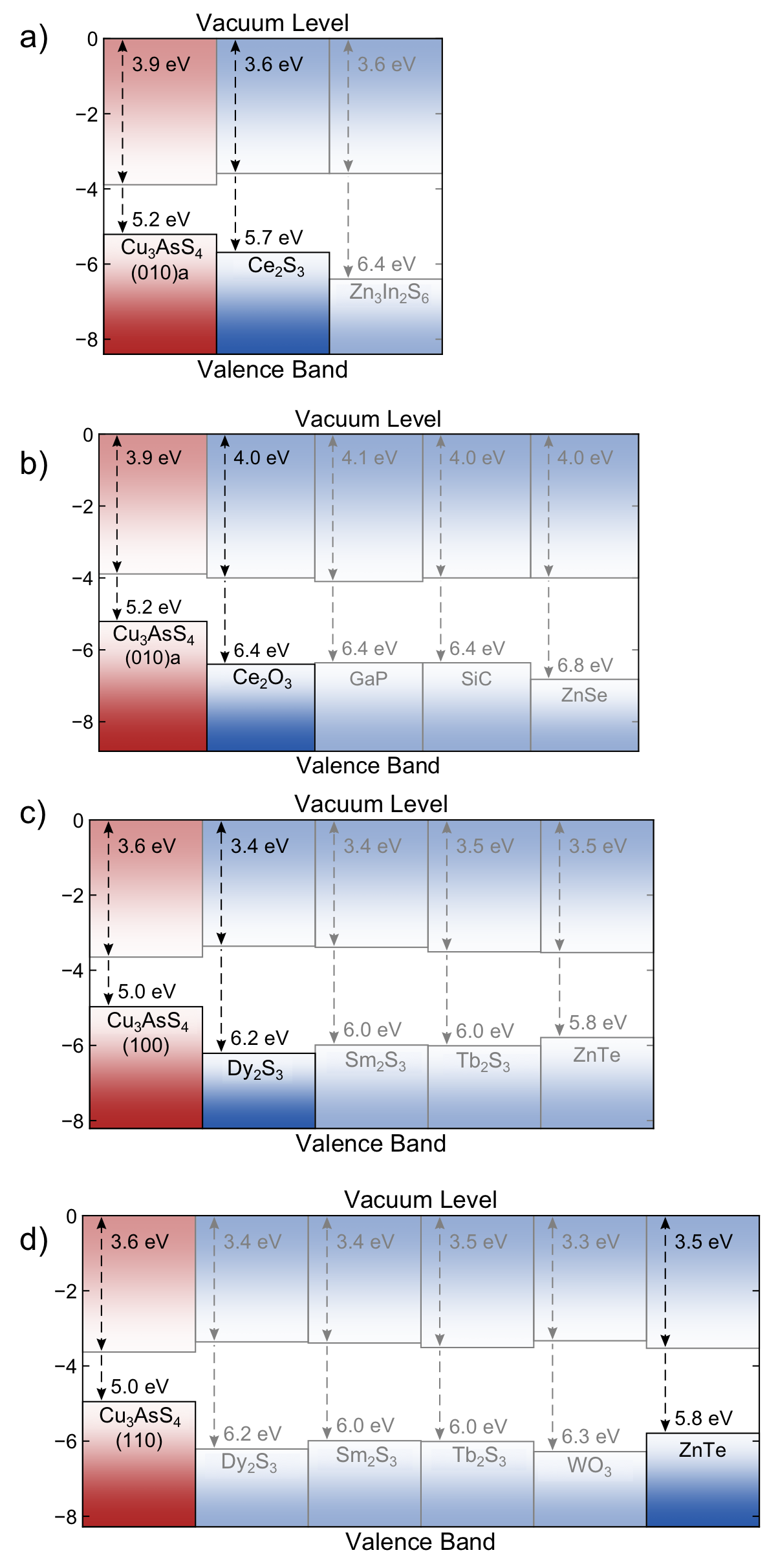}
  \caption{Candidate junction partners for p-type enargite ({\enargite}) absorber layer termination (010)a for: a) spike conduction band offset (CBO), b) cliff CBO, c) spike CBO for (100) termination and d) spike CBO for (110) termination. A low strain junction partner is highlighted in each case. 
 Band alignment plots produced using the bapt package \cite{bapt}.}
  \label{e_010a_spike+cliff}
\end{figure}

Each surface slab termination was screened individually starting from the tabulated data for 173 candidate heterojunction partners. 
The remaining candidates after the two-step screening process for each surface model are listed in Tables \ref{bournonite_jps} and \ref{enargite_jps} for bournonite and enargite, respectively. 
Averaged in-plane interface strain for all low-strain electronically-matched candidate junction partners are also reported. 
A selection of band alignment plots are shown in Figures \ref{b_110b_CBOspike+CBOcliff+VBOcliff} and \ref{e_010a_spike+cliff}; 
all other slab terminations are included in the SI. 
In each case, the candidate junction partner with the lowest interface strain is highlighted.

\begingroup
\squeezetable
\begin{table*}[]
\begin{ruledtabular}
\centering
\caption{Finding a partner for five surface terminations of bournonite ({\bournonite}).
Low strain terminations and in-plane averaged interface strain (\%) of heterojunction partners after electronic band and lattice matching for all surface models of bournonite allowing for the absorber to be either p-type (through electronic matching of conduction bands via the CBO) or n-type (through matching of the valence bands via the VBO).}
\label{bournonite_jps}
\begin{tabular}{l|lll|lll|lll}
       & \multicolumn{3}{c|}{Spike conduction band offset}  & \multicolumn{3}{c|}{Cliff conduction band offset}  & \multicolumn{3}{c}{Cliff valence band offset}  \\ \hline
 Surface      & Candidate & (\textit{hkl}) & Strain & Candidate &  (\textit{hkl}) & Strain & Candidate &  (\textit{hkl})  & Strain \\ \hline
    &        &                      &            & Ce$_2$O$_3$ & (011), (101), (110)  & 1.23\%     &                   &                      &            \\
    &        &                      &            & GaP         & (011), (101), (110)  & 1.01\%   &                   &                      &            \\
   (100) &         &                      &            & SiC         & (010), (100)         & 0.73\%   &                   &                      &            \\
    &        &                      &            & SnS$_2$     & (110)                & 0.67\%    &                   &                      &            \\
      &      &                      &            & ZnSe        & (001), (010), (100)  & 0.83\%    &                   &                      &            \\ \hline
  &  La$_2$S$_3$ & (110)              & 0.71\%    & Ce$_2$S$_3$ & (001)                & 1.71\%    & AlP               & (011), (101), (110)  & 0.71\%    \\
 (010)a & Nd$_2$S$_3$ & (001)                & 0.10\%    & Cu$_2$O     & (011), (101), (110)  & 0.81\%    & MoO$_3$           & (100)                & 0.23\%    \\
 &  Sm$_2$S$_3$ & (001)                & 2.28\%   & Gd$_2$S$_3$ & (011)                & 0.71\%    & CuI               & (110)                & 2.73\%    \\
&  WO$_3$      & (110)                & 0.49\%    & ZnTe        & (001), (010), (100)  & 0.75\%    &                   &                      &            \\ \hline
 & La$_2$S$_3$ & (110)                & 0.71\%    & Ce$_2$S$_3$ & (001)                & 1.71\%    & AlP               & (011), (101), (110)  & 0.71\% \\
(010)b & Nd$_2$S$_3$ & (001)                & 0.10\%    & Cu$_2$O     & (011), (101), (110)  & 0.81\%    & MoO$_3$           & (100)                & 0.23\%     \\
  & Sm$_2$S$_3$ & (001)                & 2.28\%    & Gd$_2$S$_3$ & (011)                & 0.71\%    & CuI               & (110)                & 2.73\%     \\
 & WO$_3$      & (110)                & 0.49\%    & ZnTe        & (001), (010), (100)  & 0.75\%    &                   &                      &            \\ \hline
& As$_2$S$_3$ & (101)                & 0.90\%    &             &                      &            & Ce$_2$O$_3$       & (011), (101), (110)  & 0.68\%    \\
& Bi$_2$O$_3$ & (100)                & 1.60\%    &             &                      &            & GaP               & (011), (101), (110)  & 1.35\%    \\
(110)a & CoTiO$_3$   & (110)                & 1.03\%    &             &                      &            & SnS$_2$           & (010), (100)         & 0.94\%    \\
 & NiTiO$_3$   & (110)                & 1.08\%    &             &                      &            & WO$_3$            & (010)                & 1.00\%    \\
  &            &                      &            &             &                      &            & Zn$_3$In$_2$S$_6$ & (110)                & 0.09\%    \\
  &           &                      &            &             &                      &            & Dy$_2$S$_3$       & (110)                & 1.47\%    \\
  &          &                      &            &             &                      &            & SiC               & (110)                & 1.47\%    \\ \hline
& GaP       &        (011), (101), (110)        &    1.35\%       &     As$_2$S$_3$        &      (101)    &   0.90\%        &      Nd$_2$S$_3$      &       (110)      &   0.64\%     \\
 &   SnS$_2$          &    (101), (100)       &     0.94\%      &    Bi$_2$O$_3$         &     (100)     &    1.60\%     &         Sm$_2$S$_3$          &         (110)        &    0.70\%    \\   
(110)b &  &  &  & CdS & (110) & 2.17\% & Tb$_2$S$_3$ & (110) & 1.71\% \\
 &  &  &  & CoTiO$_3$ & (110) & 1.03\% & ZnTe & (011), (101), (110) & 0.43\% \\
&     &  &  & PbO & (110) & 1.67\% &  &  &
\end{tabular}
\end{ruledtabular}
\end{table*}
\endgroup

\subsubsection{Junction partners for bournonite }

The lowest energy surfaces for bournonite were for the (100) and (110)b terminations, see Table \ref{elec_offsets}. 
For the bournonite (100) surface, we were only able to find low-strain partners for a CBO cliff, which is relevant for bournonite as a p-type absorber. Table \ref{bournonite_jps} and Fig. \ref{b_110b_CBOspike+CBOcliff+VBOcliff}a show several options for junction partners.
SnS$_2$ is found to be the minimum strain candidate for the (100) surface.
A further consideration is the simplicity of junction fabrication. 
Candidate junction partners able to produce a low-strain interface for multiple facets of the junction partner may imply more robust synthesis. 
For this reason, GaP and Ce$_2$O$_3$ are highlighted as alternative options.

For bournonite (110)b, we identified candidate junction partners for a spike CBO, cliff CBO, and cliff VBO. 
Candidate junction partners containing rare-earth elements such as Sm, Tb or Dy are less desirable for practical devices. 
SnS$_2$ again emerges as a promising option for a spike CBO, Bi$_2$O$_3$ for a low-toxicity cliff CBO, and ZnTe for a cliff VBO for bournonite as an n-type absorber.

Overall, for bournonite SnS$_2$ appears to be a promising junction partner for bournonite as a p-type absorber for both terminations and ZnTe for bournonite as an n-type absorber.
It is possible that there could be significant intermixing between a S containing absorber and a Te containing junction partner, as has been observed at the CdTe:CdS interface,\cite{CdTe_mixing}
which would require careful processing and may change the nature of the band offsets.

\subsubsection{Junction partners for enargite }
For enargite, the (010)a termination is the lowest energy with the next being the (100) and (110) terminations with similar energies, see Table \ref{elec_offsets}.
We note that although CdS is in our database of candidate junction partners, it did not make it through the screening process. 
This is likely a contributing factor to the low open-circuit voltage of enargite solar cells fabricated with CdS as the junction partner in Ref. \citenum{enargite_SC}. 

For the minimum energy (010)a termination of enargite, junction partners with low strain were found for both a cliff and a spike CBO, Table \ref{enargite_jps}. 
For this termination, Ce$_2$O$_3$  and Ce$_2$S$_3$ produced the lowest strain interfaces for a cliff and spike CBO, respectively.
For the (100) and (110) terminations of enargite, only low-strain candidates were found for a spike CBO but both terminations had a similar list of candidate junction partners. 
Dy$_2$S$_3$ produced the lowest strain interface with the enargite (100) surface.
However, due to the rarity of Dy, Sm and Tb, we regard candidates containing these elements as less desirable. 
ZnTe is a more promising candidate for the (100) termination. 
For the (110) surface, ZnTe is the minimum strain candidate, with WO$_3$ being another viable option. 
ZnTe is likely to be the most robust choice owing to the similar matching to the (100) and (110) surface of enargite, 
again providing that significant interface mixing or defect formation does not occur. 

\begin{table}[]
\begin{ruledtabular}
\centering
\caption{Finding a partner for five surface terminations of enargite ({\enargite}).
Identified low-strain terminations and in-plane averaged interface strain (\%) after electronic band and lattice matching.}
\label{enargite_jps}
\begin{tabular}{l|lll}
                                                              Surface              & Candidate &  (\textit{hkl})  & Strain \\ \hline
\multirow{4}{*}{\begin{tabular}[c]{@{}l@{}}(100)\\ spike CBO\end{tabular}}  &   Dy$_2$S$_3$            & (001)                          & 0.81\%    \\

& Sm$_2$S$_3$            & (001)                          & 1.43\%                         \\
& Tb$_2$S$_3$            & (001)                          & 1.10\%                           \\
& ZnTe                   & (011), (101), (110)            & 1.01\%                         \\ \hline
\multirow{2}{*}{\begin{tabular}[c]{@{}l@{}}(010)a\\ spike CBO\end{tabular}} &      Ce$_2$S$_3$            & (001)                          & 1.05\%  \\
& Zn$_3$In$_2$S$_6$      & (110)                          & 1.19\%             \\ \hline
\multirow{4}{*}{\begin{tabular}[c]{@{}l@{}}(010)a\\ cliff CBO\end{tabular}} &  Ce$_2$O$_3$     & (001), (010), (100)      & 0.20\%        \\
& GaP                & (001), (010), (100)              & 0.83\%               \\
& SiC                & (010), (100)                     & 1.33\%               \\
 & ZnSe               & (001), (010), (100)              & 2.02\%               \\ \hline
\begin{tabular}[c]{@{}l@{}}(010)b\\ spike CBO\end{tabular}      &    Bi$_2$O$_3$        & (101)         & 0.75\%        \\ \hline
\multirow{5}{*}{\begin{tabular}[c]{@{}l@{}}(110)\\ spike CBO\end{tabular}}  &     Dy$_2$S$_3$            & (011), (101)         & 0.68\% \\
& Sm$_2$S$_3$            & (001)                          & 1.15\%             \\
& Tb$_2$S$_3$            & (011), (101)                   & 0.98\%  \\
& WO$_3$                 & (011)                          & 0.33\%            \\
& ZnTe                   & (011), (101), (110)            & 0.19\%          
\end{tabular}
\end{ruledtabular}
\end{table}

\section{Conclusions}
The aim of this study has been to provide a route forward for the development of solar cell technologies based on enargite and  bournonite.
By using a combination of data-mining and first-principles calculations, we have identified promising photovoltaic heterojunction partners for low-index surface terminations of {\enargite} and  \bournonite. 
The candidate partner materials include SnS$_2$, ZnTe, WO$_3$, and Bi$_2$O$_3$.

One aspect not covered in this report is the potential photoferroic nature of these absorber materials and devices.  
Internal electric fields in polar semicondictors may suppress electron-hole recombination by enhanced local carrier separation\cite{Kirchartz_review} to provide a means of achieving high-efficiency solar cells. 
The orientation of electric polarisation may influence the direction of charge transport and collection --- as has been explored in photoanodes for photoelectrochemical water splitting applications \cite{Wang_FE}. 
These factors could determine the optimal growth and contact orientations for maximising solar conversion efficiencies. 
Further work is required in this direction both in terms of atomistic and device modelling for non-conventional photovoltaic architectures. 

\begin{acknowledgements}
We thank Ji-Sang Park and Lee Burton for useful discussions. 
This work has been supported by the EPSRC grant no. EP/L016354/1 and EP/K016288/1.
This work benefited from access to ARCHER, the UK's national high-performance computing service, which is funded by the Office of Science and Technology through EPSRC's High End Computing Programme (EP/L000202).

\textit{Data access}:
This study used the MacroDensity \cite{MacroDensity} and ElectronLatticeMatch  \cite{ELS} python libraries. 
The workflow used in this study can also be obtained from the git repository 
at \url{https://github.com/keeeto/ElectronicLatticeMatch}.
The input and output files for the DFT calculations are available from the NOMAD repository at: \url{http://dx.doi.org/10.17172/NOMAD/2018.10.25-1}.

\end{acknowledgements}

\bibliography{sulfosalts_contacts}

\end{document}